\documentclass[aps,prd,preprint,floatfix,nofootinbib,superscriptaddress]{revtex4}

\usepackage{graphicx,subfigure,epsfig,epsf,color}
\usepackage{amssymb,amsmath}
\usepackage{slashed}
\usepackage{feynmf}
\usepackage{subcaption}
\usepackage{hyperref}

\def\Lda{\Lambda}

\newcommand{\s}{\sigma}

\newcommand{\beq}{\begin{equation}}
\newcommand{\eeq}{\end{equation}}
\newcommand{\bea}{\begin{eqnarray}}
\newcommand{\eea}{\end{eqnarray}}
\newcommand{\beas}{\begin{eqnarray*}}
\newcommand{\eeas}{\end{eqnarray*}}
\newcommand{\bcr}{\begin{center}}

\def\Re{{\cal R \mskip-4mu \lower.1ex \hbox{\it e}\,}}
\def\Im{{\cal I \mskip-5mu \lower.1ex \hbox{\it m}\,}}

\def\etal{{\it et al.}}

\def\tev{\,{\ifmmode\mathrm {TeV}\else TeV\fi}}
\def\gev{\,{\ifmmode\mathrm {GeV}\else GeV\fi}}
\def\mev{\,{\ifmmode\mathrm {MeV}\else MeV\fi}}
\def\to{\rightarrow}

\begin{document}

\def\issue(#1,#2,#3){#1 (#3) #2} 
\def\APP(#1,#2,#3){Acta Phys.\ Polon.\ \issue(#1,#2,#3)}
\def\ARNPS(#1,#2,#3){Ann.\ Rev.\ Nucl.\ Part.\ Sci.\ \issue(#1,#2,#3)}
\def\CPC(#1,#2,#3){comp.\ Phys.\ comm.\ \issue(#1,#2,#3)}
\def\CIP(#1,#2,#3){comput.\ Phys.\ \issue(#1,#2,#3)}
\def\EPJC(#1,#2,#3){Eur.\ Phys.\ J.\ C\ \issue(#1,#2,#3)}
\def\EPJD(#1,#2,#3){Eur.\ Phys.\ J. Direct\ C\ \issue(#1,#2,#3)}
\def\IEEETNS(#1,#2,#3){IEEE Trans.\ Nucl.\ Sci.\ \issue(#1,#2,#3)}
\def\IJMP(#1,#2,#3){Int.\ J.\ Mod.\ Phys. \issue(#1,#2,#3)}
\def\JHEP(#1,#2,#3){J.\ High Energy Physics \issue(#1,#2,#3)}
\def\JPG(#1,#2,#3){J.\ Phys.\ G \issue(#1,#2,#3)}
\def\MPL(#1,#2,#3){Mod.\ Phys.\ Lett.\ \issue(#1,#2,#3)}
\def\NP(#1,#2,#3){Nucl.\ Phys.\ \issue(#1,#2,#3)}
\def\NIM(#1,#2,#3){Nucl.\ Instrum.\ Meth.\ \issue(#1,#2,#3)}
\def\PL(#1,#2,#3){Phys.\ Lett.\ \issue(#1,#2,#3)}
\def\PRD(#1,#2,#3){Phys.\ Rev.\ D \issue(#1,#2,#3)}
\def\PRL(#1,#2,#3){Phys.\ Rev.\ Lett.\ \issue(#1,#2,#3)}
\def\PTP(#1,#2,#3){Progs.\ Theo.\ Phys. \ \issue(#1,#2,#3)}
\def\RMP(#1,#2,#3){Rev.\ Mod.\ Phys.\ \issue(#1,#2,#3)}
\def\SJNP(#1,#2,#3){Sov.\ J. Nucl.\ Phys.\ \issue(#1,#2,#3)}

\bibliographystyle{revtex}

\title{$q$-deformed statistics and the role of a light dark matter fermion in the supernova SN1987A cooling}

\author{Atanu Guha}
\email[]{p2014401@goa.bits-pilani.ac.in}
\author{Selvaganapathy.~J}
\email[]{p2012015@goa.bits-pilani.ac.in}
\author{Prasanta~Kumar~Das}
\email[]{Author(corresponding):pdas@goa.bits-pilani.ac.in}
\affiliation{Department of Physics, Birla Institute of Technology and Science-Pilani, K. K. Birla Goa campus, NH-17B, Zuarinagar, Goa-403726, India }

\date{\today}

\begin{abstract} 
Light dark matter($\simeq 1-30~\rm{MeV}$) particles pair produced in electron-positron annihilation 
$ e^-e^+ \stackrel{\gamma}{\longrightarrow} \chi \bar{\chi}$ inside the supernova core can take away 
the energy released in the supernova SN1987A  explosion. Working within the formalism of $q$-deformed 
statistics (with the average value of the supernovae core temperature(fluctuating) being $T_{SN} = 30~\rm{MeV}$) and using the Raffelt's criterion on the emissivity for any new channel 
$\dot{\varepsilon}(e^+ e^- \to \chi \overline{\chi}) \le 10^{19}~{erg~g^{-1}s^{-1}}$, we find that as the deformation parameter $q$ changes 
from $1.0$ (undeformed scenario) to $1.1$(deformed scenario), the lower bound on the scale $\Lda$ of the dark 
matter effective theory varies from $3.3\times 10^6$ TeV to $3.2 \times 10^7$ TeV for a dark matter fermion of mass $m_\chi = 30~\rm{MeV}$. 
Using the optical depth criteria on the free streaming of the dark matter fermion, we find the lower 
bound on $\Lambda \sim 10^{8}~\rm{TeV}$ for  $m_\chi = 30~\rm{MeV}$. In a scenerio,where the dark matter 
fermions are pair produced in the outermost sector of the supernova core (with radius $0.9 R_c \le r \le R_c$, $R_c (=10~\rm{km})$ 
being the supernova core radius or the radius of proto-neutron star), we find that the bound on $\Lambda$ ($\sim 3 \times 10^7$ TeV)
obtained from SN cooling criteria (Raffelt's criteria) is comparable with the bound  obtained from free streaming (optical depth criterion) for light fermion dark matter of mass 
 $m_{\chi}=10 - 30$ MeV.



\end{abstract}

\maketitle

\section{Introduction}

The fact that dark matter plays a very important role in building the universe we live in is gradually gaining
ground due to concrete experimental evidence collected over a period of time.  
In 1933, Fritz Zwicky \cite{Zwicky} found that the normal luminous matter is alone not sufficient 
to explain the  velocity dispersion of galaxies in the Coma cluster of galaxies: one requires non-luminous 
matter, dubbed as dark matter(DM). Current data suggests that DM is five times more than the normal luminous 
matter in the Universe\cite{Feng}.   Now, the dark matter(DM) does not interact electromagnetically with the normal luminous 
matter since it(DM) has no electromagnetic charge. Even it does so, it is 
very weak. So far, scientists have been able to infer the existence of dark matter 
only through its gravitational effect on normal matter. 

But what is dark matter? For a long time it remains a mystery. A wide range of collider and astrophysical study suggests that it is 
a Weakly Interacting Massive Particle(WIMP) of mass lying in between few MeV to few GeV. Theories suggest that DM 
candidates are most likely to be found in the beyond the Standard Model(SM) physics, such as supersymmetry and 
extra dimensions. Direct detection of DM includes its interaction with nucleons in underground detectors,
whereas indirect detection through DM annihilation to SM states in the Sun (to neutrinos) has been done.
Experiments at the Large Hadron Collider(LHC) and the 
upcoming electron-positron linear collider(LC) will give more information about the dark matter as the  
missing energy signature. See \cite{JKG,BHS,Feng} for a review on dark matter searches.

 An enormous amount of gravitational binding energy $10^{53}~\rm{ergs}$  was released in the supernova SN1987A 
 explosion of which about $99\%$ was carried away by neutrino alone. To understand the supernova energy loss mechanism 
 and its relevance in the beyond standard model physics has been an area of 
 active research for a long time\cite{Raffelt}. In 2006, Fayet \etal \cite{FHS} studied the impact of light dark matter on the core 
 collapse supernova cooling and found that the  $1 - 30~\rm{MeV}$ mass dark matter fermion can explain SN1987A energy loss rate. They also found that if the dark matter 
 particles are of mass $m_\chi \le 10~\rm{MeV}$ and reproduce 
the observed dark matter relic density, it would lead to the modification of the supernova cooling dynamics, 
which is unacceptable.  Kadota \etal \cite{KS} studied the impact of International Linear Collider(ILC) and SN1987A energy loss rate due to the 
light (MeV) dark matter. They found the SN bound to be more stringent than those obtained from ILC by a 
factor of $\mathcal{O}(10^5)$ for a DM mass below $100$ MeV.  
 
 Here we would like to investigate the impact of light dark matter fermions on the energy released in SN1987A explosion. We work in a 
 dark matter model characterized by an effective scale $\Lda$ and use 
the formalism of $q$-deformed statistics which takes care of the fluctuation of the core temperature of the 
supernova. While Kadota \etal \cite{KS} considered the dark matter fermion and SM photon coupling 
arising from magnetic dipole moment operator, we have generalized our work where a SM photon may couple 
with dark matter fermion through the magnetic dipole moment operator or electric dipole moment operators
or both.  
      
~  The outline of the work is as follows. We give a brief description of SN1987A 
cooling problem and a small introduction of $q$-defomed statistics in section II. In section III, we discuss the dark 
matter pair production in electron-positron annihilation and find the supernova energy loss rate due 
to this dark matter fermion pair production. The numerical analysis part is presented in section IV. 
Using the Raffelt's criterion, we obtain bound on the scale $\Lda$ of the effective dark matter theory in 
deformed($q\neq 1$) and undeformed ($q=1$) scenarios. Using the optical depth criteria(based on free streaming 
of dark matter fermions), we obtain constraints on $\Lda$. Finally, we conclude in section V and VI.

\section{Supernova Explosion, its Cooling and $q$-deformed statistics}
\subsection{Supernova cooling, Raffelt's criterion }

 The supernova SN1987A, a typical example of a core-collapse supernova explosion is the final fate 
 of massive star of mass  $M \ge 8\,M_\odot$. The energy released in SN1987A explosion is 
 enormous: it is the gravitational binding energy $E_g$ of the proto-neutron star (of mass $M_{PNS}$), given by
\beq
E_g = \frac{3 G_N M_{PNS}^2}{5 R_{NS}} \sim 3.0 \times 10^{53} \;{\rm erg.}
\eeq
Here $M_{PNS} = 1.5 M_\odot$, $R_{NS} = 10~{\rm Km}$ and $G_N$ is the Newton's gravitational constant. 
Out of this $99\%$ of the released energy is carried away by neutrinos, while the rest $1\%$ contributes to the kinetic energy of the explosion. To detect this neutrino burst by the 
earth based detector is of primary astrophysical interest of the core-collapse supernova. About 
$10^{53}~\rm{ergs}$ energy was released in the supernova SN1987A explosion 
in a couple of seconds and two collaborations Kamiokande \cite{Kamioka} and IMB\cite{IMB}
first detected this neutrino flux using their earth based detectors. 
The observed neutrino luminosity in the detector(IMB or Kamiokande) 
is $L_\nu \sim 3 \times 10^{53}\;{\rm erg ~s^{-1}}$ (including $3$ generations of neutrinos and anti-neutrinos 
i.e. $\nu_e,~\nu_\mu,~\nu_\tau$ and ${\overline{\nu_e}},~{\overline{\nu_\mu}},~{\overline{\nu_\tau}}$). 
So $\tilde{L}_\nu = \frac{L_\nu}{6} \sim 3 \times 10^{52}~{\rm erg~ s^{-1}}$. The mass of a typical 
proto-neutron star $M_{PNS} = 1.5 M_\odot = 3 \times 10^{33}~{\rm g}$. So, the average energy loss per unit mass is $\frac{\tilde{L}_\nu}{M_{PNS}} \simeq 1 \times 10^{19}~{\rm erg ~g^{-1}s^{-1} }$. 
Note that this is the energy carried away by each of the above 6 (anti)-neutrino species. 
Now besides neutrino, if KK graviton, KK radion, axion also take away energy, 
the energy-loss rate due to these new channels $\epsilon_{new}$ should be less than the above average energy 
loss rate \cite{Raffelt} i.e. 
\beq
\epsilon_{new} \le 10^{19}~{\rm erg ~g^{-1}s^{-1} }
\eeq
and this follows from the observed neutrino luminosity per species (total $6$ neutrino and anti-neutrinos, three type each). So, the upper bound on $\epsilon_{new}$ (G. Raffelt's criterion) is a data-driven entity and it does not have an implicit dependence on the fact the ensemble of particles obey the normal Bose-Einstein(BE) statistics or Fermi-Dirac(FD) Statistics characterized by the equilibrium temperature $T_{SN}$ or by Tsallis statistics where the fluctuation around $T_{SN}$ is taken into account. 
If any energy-loss mechanism has an emissivity greater than $10^{19}~\rm{erg~g^{-1}~s^{-1}}$, then it will remove 
sufficient energy from the explosion to invalidate the current understanding of core-collapse supernova.

Using the Raffelt's criteria of  the supernova energy loss rate for any new physics channel, we now 
constrain the scale $\Lda$ of the dark matter effective theory. Now since the core temperature of the supernova is fluctuating,  we will work here 
within the formalism of $q$-deformed statistics\cite{Tsallis} where this temperature fluctuation is taken into 
account.

\subsection{Fluctuating temperature and Tsallis statistics and temperature fluctuation}
The temperature ($T$) fluctuations  in the $q$-deformed statistics \cite{Beck_cohen} takes the 
$\chi^2$ distribution  of the following form  
\bea
f(\beta) = \frac{1}{\Gamma\left(\frac{n}{2}\right)} \left(\frac{n}{2 \beta_0} \right)^{n/2} \beta^{\frac{n}{2} - 1} ~exp\left(- \frac{n \beta}{2 \beta_0}\right)
\label{fbeta}
\eea 
where $n$ is the degree of the distribution and $\beta = {1 \over {k T}}$. 
The average of the fluctuating inverse temperature $\beta$ can be estimated as 
\bea
\langle \beta \rangle = n \langle X_i^2 \rangle = \int_{0}^{\infty} \beta f(\beta) d\beta = \beta_0
\eea 
Taking into account the local temperature fluctuation, integrating over all $\beta$, we find
the $q$-generalized relativistic( with particle energy $E = \sqrt{{\bf p}^2 c^2 + m^2 c^4}$) 
Maxwell-Boltzmann distribution 
\bea
{\mathcal{P}}(E) \sim  \frac{E^2}{\left(1 + b(q-1)E\right)^{\frac{1}{q-1}}}
\eea
where  $q = 1 + \frac{2}{n + 6}$ and $b = \frac{\beta_0}{4 - 3 q}$. It's generalization to Fermi-Dirac 
and Bose-Einstein distribution is worked out in \cite{Beck_super}.  The average occupation number of 
any particle within this $q$-deformed statistics ( Tsallis statistics \cite{Tsallis}) formalism, 
is given by  $f_i(\beta,E_i)$ ($i = 1,~2$ corresponds to particles) where
\bea
f_i(\beta,E_i) = \frac{1}{\left(1 + (q-1)b E_i \right)^{\frac{1}{q-1}} \pm 1}
\eea
where the $-$ sign is for bosons and $+$ sign is for fermions. 
Note that the effective Boltzmann factor 
$x_i = \left(1 + (q-1)b E_i \right)^{-\frac{1}{q-1}}$ approaches to the ordinary Boltzmann factor 
$e^{- b E_i} (= e^{- \beta_0 E_i})$ as $q \to 1$.   
The $q$-deformed statistics finds important application in collider physics and astrophysics: 
Beck \etal \cite{Beck_cosmic} use the $q$-deformed statistics in order to explain the 
measured energy spectrum of primary cosmic rays. With $b^{-1} = k T_0 = 107~\rm{MeV}$ and the 
deformation parameter $q = 1.215$  and $q = 1.222$,respectively, they were able to explain quite well 
the flux rate i.e. the upper(upto the knee) portion and the lower (the ankle) portion of 
the cosmic ray spectrum \cite{Beck_super}.  Bediaga \etal 
use the $q$-deformed statistics to explain the differential cross-section for transverse momenta in electron-positron 
annihilation \cite{Bediaga-ep}.  The applications of $q$-deformed statistics 
for chaotically  quantized scalar fields \cite{Beck_deform}, dark energy \cite{Beck_dark} are available in the literature.
Recently, Das \etal  \cite{Das} found that an ultra-light 
radion in brane-world Randall-Sundrum model produced in the supernova core can also take 
away energy released in SN1987A explosion and thus can explain the SN1987A energy loss rate provided 
$q$ lies within the range $1.18 < q < 1.32$. Here we investigate the supernova SN1987A cooling due to free streaming of 
fermionic dark matter.

\section{Dark matter pair production inside supernova}
Electrons are abundant in supernova. The dark matter fermion may be pair produced in the $s$-channel 
annihilation of electron and positron: $ e^-(p_1)e^+(p_2) \stackrel{\gamma}{\longrightarrow} \chi(p_3)
\bar{\chi}(p_4) $.
\begin{figure}[htbp]
\includegraphics[width=7cm]{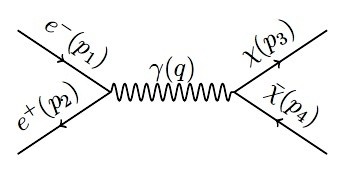}
\vspace*{-0.15in}
\caption{{\it  Feynman diagram for the process $ e^-e^+ \stackrel{\gamma}{\longrightarrow} \chi
\bar{\chi}$ }}
\end{figure}

The effective Lagrangian of describing photon($\gamma$) and dark matter fermion ($\chi $) interaction is given 
by
\bea 
\mathcal{L}=-\frac{i}{2}\bar{\chi}\sigma_{\mu \nu}(\mu_{\chi}+\gamma_5 d_{\chi})\chi F^{\mu \nu}
\eea
\noindent where $F^{\mu\nu} = \partial^\mu A^\nu - \partial^\nu A^\mu$, the e.m.field strength tensor. 
Here $ \mu_{\chi} $ and $ d_{\chi} $ correspond to the magnetic dipole moment and the electric dipole 
moment  of the dark matter fermion $\chi$. $\sigma^{\mu \nu}=\frac{i}{2}[\gamma^\mu,\gamma^\nu] $ is 
the spin tensor. The $4$-momentum vectors of the initial and final state particles in the centre-of-mass
frame are given by
\begin{eqnarray}
p_{1} &=&  \left(E, 0, 0, p_z \right); ~p_{2} = \left(E, 0, 0,-p_z \right); \nonumber \\  
p_{3} &=& \left(E^\prime, p {\rm sin}\theta {\rm cos}\phi, p {\rm  sin}\theta {\rm sin}\phi, p {\rm cos}\theta \right);\nonumber \\
p_{4} &=& \left(E^\prime,-p {\rm sin}\theta {\rm cos}\phi,- p {\rm sin}\theta {\rm sin}\phi,-p {\rm cos}\theta \right). \nonumber 
\end{eqnarray}  
The spin-averaged amplitude-square for the process $ e^-(p_1)e^+(p_2) \stackrel{\gamma}{\longrightarrow} \chi(p_3)
\bar{\chi}(p_4) $ is given by 
\bea \label{eqn:Spin averaged Amplitude Square}
\overline{|\mathcal{M}|^2}=4 \pi \alpha[\mu_\chi^2\left\lbrace s (1-\cos^2 \theta)+4m_\chi^2(1+\cos^2 \theta)\right\rbrace+d_\chi^2\left\lbrace (s-4m_\chi^2) (1-\cos^2 \theta)\right\rbrace]
\eea
\noindent 

The differential cross-section for the process is 
\bea \label{eqn:dsigmadOmega}
\frac{d\sigma}{d\Omega}(e^-e^+ \stackrel{\gamma}{\longrightarrow} \chi
\bar{\chi})=\frac{1}{64\pi^2 s}\cdot\sqrt{1-\frac{4m_\chi^2}{s}}\cdot\overline{|\mathcal{M}|^2}
\eea
\noindent 


Finally, the total cross-section is given by 
\bea \label{eqn:totalcross-section}
\sigma(e^-e^+ \stackrel{\gamma}{\longrightarrow} \chi
\bar{\chi})=\frac{\alpha}{6s} \cdot \sqrt{1-\frac{4m_\chi^2}{s}} \cdot \left[\mu_\chi^2(s+8m_\chi^2) + d_\chi^2 (s-4m_\chi^2) \right]
\eea
\noindent 
Here $m_\chi$ is the dark matter mass, $\alpha = \frac{e^2}{4 \pi}$ and  
$s=(p_1 + p_2)^2 = (p_3 + p_4)^2$ is the Mandelstam variable.
\newpage
\begin{flushleft}
{\bf Energy Loss rate} 
\end{flushleft}
The supernova energy loss rate due to dark matter fermion pair production  is given by \cite{Raffelt}
\bea \label{eqn:energy-loss-rate}
\dot{\varepsilon}_{e^- e^+ \rightarrow \chi \overline{\chi}} 
& = & \frac{1}{\rho_{SN}} \langle n_{e^-} n_{e^+} \sigma_{e^- e^+ \rightarrow \chi \overline{\chi}}~ V_{rel} E_{com}\rangle \nonumber \\
 &=& \frac{1}{\rho_{SN}} \frac{1}{\pi^4} \int_{m_\chi}^{\infty} \int_{m_\chi}^{\infty} d E_1 d E_2  
\frac{E_1 E_2 (E_1 + E_2)^3}{2 D_1 D_2}~\s_{e^- e^+ \rightarrow \chi \overline{\chi}} 
\eea 
where the c.o.m energy $ E_{com}(=E_1+E_2) =2 E$ (where $E_1 = E_2 = E$) and the relative velocity 
$ V_{rel}=\frac{s}{4E_1 E_2}$. $\rho_{SN}$ is the supernova matter density.  The cross-section $\sigma_{e^- e^+ \rightarrow \chi \overline{\chi}}$ is given in Eq.~\ref{eqn:totalcross-section} and  
$D_i = \left(1 + \frac{b}{\tau} (E_i - \mu_i) \right)^{\tau} + 1$ with $\rm{i} = 1,2$. Here 
$b=\frac{\beta_0}{4 - 3q}$, $\beta_0 = \frac{1}{k_B T}$ (we are working in the unit where $k_B=1$) and 
$\tau = {1 \over q-1}$. The electron and positron number densities 
$n_{e^-}= \int \frac{2 d^3p_1}{(2\pi)^3}~{D_1}^{-1}$ and $n_{e^+}= \int \frac{2 d^3p_2}{(2\pi)^3}~{D_2}^{-1}$.
Introducing the dimensionless variables $x_i = E_i/T$~($i=1,2$), we can finally write the energy loss rate
Eq.~\ref{eqn:energy-loss-rate} as, 
\bea \label{eqn:energy loss rate final}
\dot{\varepsilon}_{e^- e^+ \to \chi \overline{\chi}}=\frac{\alpha T^{7}}{12\pi^4\rho_{SN}} \int_{\frac{m_\chi}{T}}^{\infty} dx_1 
\int_{\frac{m_\chi}{T}}^{\infty} dx_2 ~\frac{x_1~x_2~(x_1+x_2)}
{\left[\left(1 + \frac{b}{\tau} (T x_1 - \mu_{e^-}) \right)^{\tau} + 1\right]}~
\frac{\mathcal{F}}{\left[\left(1 + \frac{b}{\tau} (T x_2 - \mu_{e^+})\right)^{\tau} + 1\right]} \nonumber \\
\eea
\noindent 
where the function $\mathcal{F}$ is given by
\bea \label{eqn:X}
{\mathcal F}=\sqrt{1-\frac{4m_\chi^2}{T^2(x_1+x_2)^2}} \cdot \left[\mu_\chi^2\left\lbrace(x_1+x_2)^2+\frac{8m_\chi^2}{T^2}\right\rbrace + d_\chi^2 \left\lbrace(x_1+x_2)^2-\frac{4m_\chi^2}{T^2}\right\rbrace \right]
\eea
\noindent 

Noting the fact that in the $q \to 1$ limit, the $q$-deformed distribution formula gets converted to 
either the  Bose-Einstein or Fermi-Dirac statistical distribution formula (which describes the un-deformed scenario)
(see the APPENDIX for a proof) i.e. 
\bea
f_i(\beta,E_i) = \frac{1}{\left(1 + (q-1)b E_i \right)^{\frac{1}{q-1}} \pm 1} ~\stackrel{q \to 1}{\longrightarrow} \frac{1}{e^{b E_i} \pm 1} \left(= \frac{1}{e^{\beta_0 E_i} \pm 1}\right)
\eea
where $e^{b E_i} = e^{\beta_0 E_i} $ with $b = \frac{\beta_0}{4 - 3 q} = \beta_0$ for $q \to 1$ and 
$\beta_0$ is the inverse equilibrium temperature $T_0$ of the supernova core, the energy loss rate 
in $q=1$ case takes the following form 
\bea \label{eqn:energy-loss-rate-undeformed}
\dot{\varepsilon}_{e^- e^+ \to \chi \overline{\chi}}=\frac{\alpha T^{7}}{12\pi^4\rho_{SN}} 
\int_{\frac{m_\chi}{T}}^{\infty} dx_1 \int_{\frac{m_\chi}{T}}^{\infty} dx_2 ~\frac{x_1(x_1+x_2)}
{\left[\exp {\left(x_1- \frac{\mu_{e^-}}{T} \right)}  + 1\right]}~\frac{x_2}
{\left[\exp {\left(x_2- \frac{\mu_{e^+}}{T} \right)}  + 1\right]}  ~\mathcal{F}
\eea 
where, $\mu_{e^+}=-\mu_{e^-}$

\section{Numerical Analysis}
The dark matter produced inside the supernova core via the channel 
$e^- e^+ \to \chi \overline{\chi} $ can contribute to the supernova energy loss rate, if the 
emissivity of this channel $\dot{\varepsilon}(e^- e^+ \to \chi \overline{\chi}) \le 10^{19}~{erg~g^{-1}s^{-1}}~(=7.288\times 10^{-27} \rm{GeV})$.

Since the core temperature(T) of the supernova is fluctuating, we follow the $\chi^2$ distribution 
analysis technique \cite{Beck_cohen} here, where the temperature distribution is characterized by 
it's mean value  $T (=T_{SN}) = 30~\rm{MeV}$ (see Section II B for more details about $\chi^2$ distribution).
  Because of this temperature fluctuation, the ensemble of nucleons, electrons, 
dark matter fermions, photons inside the supernova follow a statistics popularly known as the $q$-deformed 
statistics (or Tsallis statistics \cite{Tsallis}, see section II for the related discussion), 
which is different than the usual Fermi-Dirac and Bose-Einstein statistics. The parameter 
$q$ characterizing such distribution is called the deformation parameter: for 
$q \neq 1$, it is the $q$-deformed distribution , while for $q=1$, it is the regular Fermi-Dirac or Bose-Einstein distribution
(undeformed distribution). We will investigate here the dependence of the deformation parameter $q$ on the scale $\Lda$ of 
the dark matter effective theory for a dark matter fermion of mass lying between $1-100~\rm{MeV}$.

\subsection{Bound on the effective scale $\Lambda$ from the $e^{+} + e^{-} \stackrel{\gamma}{\longrightarrow} \chi \overline{\chi}$ process}
Depending on whether the effective coupling of dark matter fermion with photon 
is characterized by a dipole moment operator of magnetic or  electric type, we have the below mentioned 
three cases:  
\begin{enumerate}
 \item Case I: $\mu_\chi (\sim 1/\Lambda_\mu) \neq 0,~~d_\chi (\sim 1/\Lambda_d) = 0$.
 \item Case II: $\mu_\chi (\sim 1/\Lambda_\mu) = 0,~~d_\chi (\sim 1/\Lambda_d) \neq 0$.
 \item Case III: $\mu_\chi (\sim 1/\Lambda_\mu) \neq 0,~~d_\chi (\sim 1/\Lambda_d) \neq 0$. 
Here $\Lda_\mu = \Lda_d = \Lda$
\end{enumerate}
In each case we have two possible scenarios: (i) Scenario I: The deformation 
parameter $q > 1$ which is known as the $q$-deformed scenario and (ii) Scenario II: The deformation 
parameter $q=1$ which is the undeformed scenario and particles obey the usual Fermi-Dirac or Bose-Einstein statistics.

\subsubsection{Scenario I: $q$-deformed statistics ($q > 1$)}
\begin{figure}[htbp]
\includegraphics[width=8.0cm]{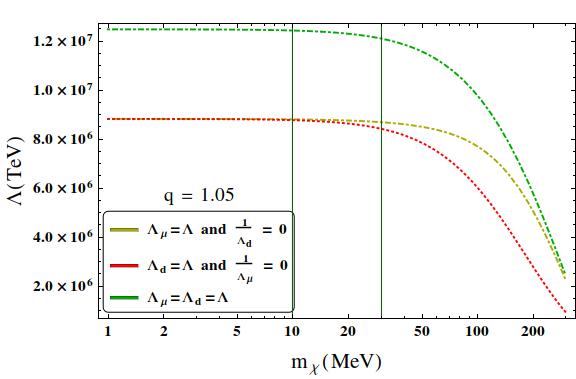} \hspace*{0.02in} \includegraphics[width=8.0cm]{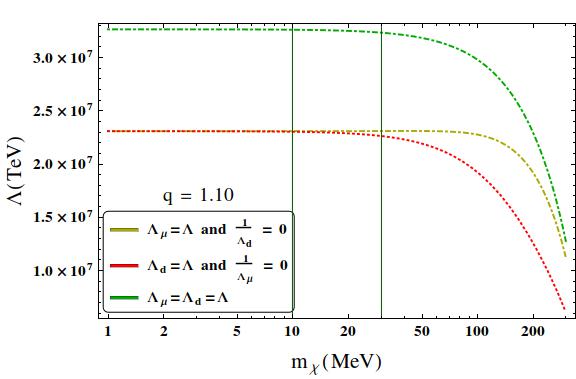}
\caption{\it $\Lda_{\mu,d}$ (in TeV) (lower two curves) and $\Lda$ (in TeV) (topmost curve) are plotted against 
$m_\chi$ (in MeV) for  $q=1.05$(left figure) and $q=1.1$(right figure), respectively.}
\protect\label{Lda-mdark-q1pt05pt1}
\end{figure}

 Using the Raffelt's criteria $\dot{\varepsilon}(e^- e^+ \to \chi \overline{\chi}) \le 10^{19}~{erg~g^{-1}s^{-1}}~(=7.288\times 10^{-27} \rm{GeV})$ and 
equation (\ref{eqn:energy loss rate final}), we derive a lower bound on the scale $\Lambda$ 
(which we denote as $\Lda_{\mu}$ and $\Lda_{d}$ in Case I and Case II, respectively) of the dark matter effective theory. 
In Fig. \ref{Lda-mdark-q1pt05pt1}, we have plotted $\Lda_{\mu},\Lda_{d}$ (Case I, Case II) and $\Lda$ 
(Case III) as a function of the dark matter fermion mass $m_\chi$ corresponding to $q=1.05$ 
(left figure) and $q=1.1$ (right figure), respectively.
The region lying on the left of the vertical line corresponding to $m_\chi = 30~\rm{MeV}$ and above the 
horizontal curves of each figure is allowed. For $10~\rm{MeV} \le m_\chi \le 30~\rm{MeV}$ 
\cite{FHS}, the effective scale $\Lda$ corresponding to that window (and above the curves) is 
allowed. 
Note that $\Lda_{\mu,d}$ remains constant (in both figures) till $m_\chi = 10~\rm{MeV}$
and starts decreasing after that. On the left figure ($q=1.05$), till 
$m_\chi = 10~\rm{MeV}$ we find $\Lda_\mu = \Lda_d = 8.8 \times 10^6$ TeV (Case I and Case II) and
$\Lda = 1.24 \times 10^7$ TeV (Case III). It becomes $\Lda_\mu =  8.7 \times 10^6$ TeV and 
$\Lda_d = 8.4 \times 10^6$ TeV (in Case I and Case II) and $\Lda = 1.2 \times 10^7$ TeV 
(in Case III) at $m_\chi = 30~\rm{MeV}$. As we go from left to right figure, $\Lda_{\mu,d}$ gets increased at a given $m_\chi$. 
For $m_\chi = 30~\rm{MeV}$, the topmost curve gives $\Lda = 3.23 \times 10^7$ TeV, whereas 
the lower two curves give $\Lda_\mu = 2.3 \times 10^7$ TeV and $\Lda_d = 2.26 \times 10^7$ TeV, 
respectively.
 
\begin{figure}[htbp]
\includegraphics[width=8cm]{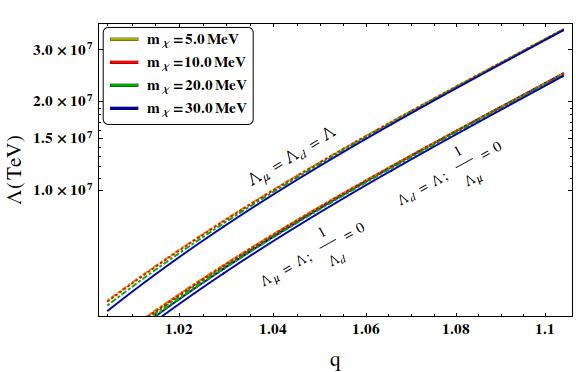}\hspace*{0.02in} \includegraphics[width=8.0cm]{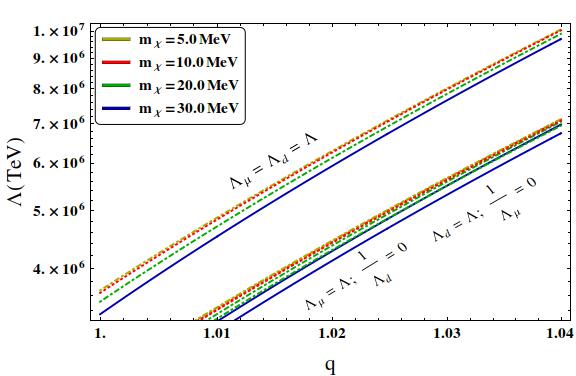}
\vspace*{-0.1in}
\caption{\it In the left figure, $\Lda_{\mu,d}$ (in TeV) (lower curves) and $\Lda$ (in TeV) 
(upper curves) are plotted as a function of $q$ for $m_\chi=5,~10,~20$ 
and $30~\rm{MeV}$. On the right, we have plotted the same but for $1.04 < q < 1$.}
\protect\label{Lda-qparam}
\end{figure}

 In Fig. \ref{Lda-qparam}  we have shown $\Lda$ as a function of the deformation parameter $q$ 
 corresponding to  $m_\chi = 5,~10,~20$ and $30$ MeV for the above three cases (using Raffelt's criteria and analyzing the equation (\ref{eqn:energy loss rate final})).  
On the right figure, the same plots are shown corresponding to $1.0 \le q \le 1.04$. 
The following observations are in order:
\begin{enumerate}
 \item For a given $m_\chi$, the lower bound on $\Lda$ increases as $q$ increases. For example, for $m_\chi = 30~\rm{MeV}$,
  $\Lda$ changes from $4.8\times 10^6~\rm{TeV}$ to $3.4\times 10^7~\rm{TeV}$ as $q$ changes from $1.01$ 
 to $1.1$ (left figure).   
 \item For a given $q$, $\Lambda$ decreases with the increase in $m_\chi$. For example, 
 for $q=1.03$, we see that as $m_\chi$ increases from $10~\rm{MeV}$ to $30~\rm{MeV}$, $\Lda$ 
 decreases from $8 \times 10^6~\rm{TeV}$ to $7.7\times 10^6~\rm{TeV}$.
\end{enumerate}


\subsubsection{Scenario II: Undeformed statistics($q=1$)} 
The q-deformed distribution becomes the usual(undeformed) Bose-Einstein or Fermi-Dirac 
type distribution for $q=1$(in Appendix A, we have given a derivation of this). We are now to investigate 
how the bound on $\Lda$ gets changed as one switches from a deformed distribution to an 
un-deformed distribution.   
\begin{figure}[htbp]
\includegraphics[width=10cm]{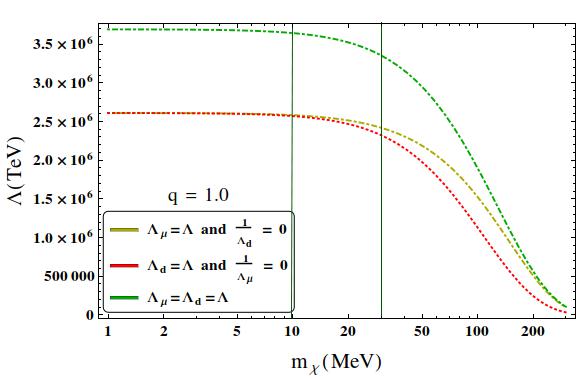}
\caption{{\it  $\Lda_{\mu,d}$(in TeV) and  $\Lda$(in TeV) are plotted against $m_\chi$ 
(in MeV) for $q=1$. }}
\protect\label{Lda-mdark}
\end{figure}
In Fig. \ref{Lda-mdark}, we have plotted $\Lda_{\mu,d}$ and $\Lda$ against $m_\chi$ for $q=1$ using the Raffelt's 
criteria and equation (\ref{eqn:energy-loss-rate-undeformed}). 
The region on the left of the vertical line at $m_\chi = 30~\rm{MeV}$ and above the horizontal 
curves is allowed. Up to $m_\chi = 10~\rm{MeV}$, we find $\Lda_\mu$ (Case I) and $\Lda_d$ (Case II) are found
to be constant at  $\sim 2.6 \times 10^6~\rm{TeV}$ and after which they start decreasing.  
The topmost curve corresponds to $\Lda$(Case~III) for 
different values of $m_\chi$. For example, corresponding to $m_\chi = 1 - 10~\rm{MeV}$, we find 
$\Lda \sim 3.6 \times 10^6~\rm{TeV}$, while at $m_\chi = 30~\rm{MeV}$, we find $\Lda = 3.3 \times 10^6~\rm{TeV}$. 
We have summarized our result in Table 1. From Table 1, we see that as $m_\chi$ increases, the bound on 
$\Lda_{\mu,d}$ and $\Lda$ decreases. 
\begin{center}
Table 1
\end{center}
\vspace*{-0.25in}
\begin{center}
\begin{tabular}{|c|c|c|c|}
\hline
$m_\chi~(\rm{MeV})$ & Case I~: $\Lda_\mu~\rm{(TeV)}$ & Case II~: $\Lda_d~\rm{(TeV)}$ & Case III~: $\Lda~\rm{(TeV)}$  \\
\hline
 $10$ & $ 2.6\times 10^6$ & $2.6\times 10^6$ & $3.6\times 10^6$ \\ 
\hline
 $30$ & $2.4\times 10^6$ & $2.3\times 10^6$ & $3.3\times 10^6$ \\
 \hline
\end{tabular}
\end{center}
\noindent {\it Table 1: The lower bound on the effective scale $\Lda_{\mu}$, $\Lda_{d}$ and $\Lda$ (TeV) are shown for different 
dark matter mass $m_\chi$ (MeV) in the undeformed scenario.}

The lower bound obtained on $\Lambda_\mu$ is comparable with that obtained by Kadota 
\etal \cite{KS}.  


\section{Dark particle Free streaming/Trapping}
The constraint on $\Lambda$ obtained in earlier section 
holds to be true if the produced dark matter fermion free streams out of the supernova.
To find the free streaming let us calculate their mean free path \cite{Dreiner}
\bea
\lambda_{\chi}=\frac{1}{n_e \cdot \sigma_{e \chi \rightarrow e \chi}}
\protect\label{mean_free_path}
\eea
where $n_e(= 8.7 \times 10^{43} m^{-3})$ is the number density of the colliding electrons in the supernova 
and $ \sigma_{e \chi \rightarrow e \chi}$ is the cross section for the scattering of 
dark matter fermion on electron which is related via the crossing symmetry to the annihilation 
cross-section $\sigma_{e e  \rightarrow \chi  \chi}$.  
Now, most of the dark matter particles produced in the outermost $ 10\% $ of the star ($ 0.9 R_{c} < r < R_{c} $) from 
electron-positron annihilation  \cite{DHLP}. Then any of the dark matter 
particles produced in electron-positron annihilation while propagating through the proto-neutron star, 
can undergo scattering due the presence of neutrons and electrons inside the star. 
In the case of supernova cooling, neutron-dark matter particle scattering will be negligible 
for free streaming due to neutron mass \cite{Dreiner}. We use the optical depth criteria \cite{Dreiner} 
\bea
\int_{r_0}^{R_c}\frac{dr}{\lambda_{\chi}} \leq \frac{2}{3}
\eea
to investigate whether the dark matter fermion produced at a depth $r_0$ free streams out of the supernova 
and takes away the released energy or getting trapped inside the supernova. Here we set 
$r_0=0.9 R_c$ in our analysis, where $R_c $($ \simeq 10 $ km) is the radius of the supernova core 
(proto-neutron star) \cite{Dreiner}. From the optical depth criteria, we find that 
minimum length of the mean free path for free streaming $\lambda_{fs}$ is $\lambda^{min}_{fs} = 1.5~\rm{km}$ 
and it increases with $\Lambda$. 
\begin{figure}[htbp]
\includegraphics[width=8cm]{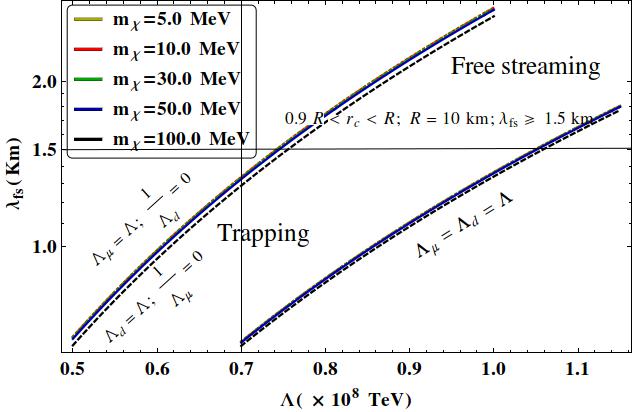}\hspace*{0.02in} \includegraphics[width=8.0cm]{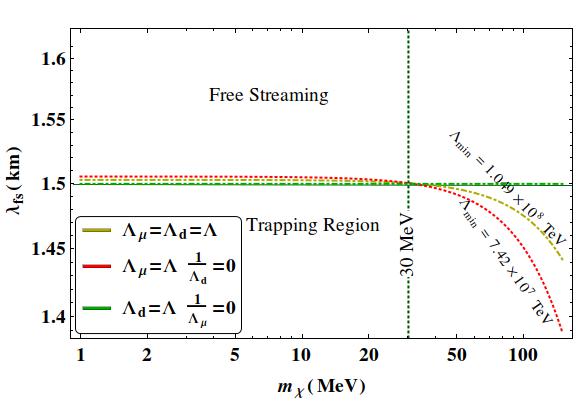}
\caption{On the left side, the free streaming length $\lambda_{fs}$ (km) is plotted as a function of $\Lambda$ (GeV) for different $m_\chi$. 
We have taken $R_c(=R) = 10~\rm{km}$. On the right side, the free streaming length $\lambda_{fs}$ (km) is plotted as a function of 
$m_\chi$ (MeV). The trapping and free streaming regions are shown separately.}
\protect\label{fig:Figure5}
\end{figure}
In Fig. \ref{fig:Figure5}(on the left side), we have 
shown the free streaming length $\lambda_{fs}$ (in km) as a function of $\Lambda$ for different  
$m_\chi$. Results for three different cases: Case I: $\Lambda_\mu = \Lambda,~\frac{1}{\Lambda_d}=0$ and 
Case II: $\Lambda_d = \Lambda,~\frac{1}{\Lambda_\mu}=0$ and Case III: $\Lambda_\mu = \Lambda_d = \Lambda$ are shown. 
The horizontal line correspond to $\lambda^{min}_{fs} = 1.5~\rm{km}$ and it's intersection with 
the curves gives the lower bound on $\Lambda_{\mu}(\Lambda_{d})=7.42 \times 10^7 ~\rm{TeV}$, in Case I (Case II) and 
$ \Lda = 1.05 \times 10^{8}~\rm{TeV}$ in Case III, respectively.  On the right side (of Figure \ref{fig:Figure5}), 
we have plotted the free streaming length $\lambda_{fs}$ against $m_\chi$. 
The horizontal lines corresponding to $\lambda^{min}_{fs} = 1.5~\rm{km}$
remain constant until $m_\chi = 30~\rm{MeV}$, after which they start 
decreasing except the Case-II in which it remains to be constant beyond $m_{\chi} = 30~\rm{MeV}$. 
The region below the line $\lambda_{fs} = 1.5~\rm{km}$ corresponds to the trapping region, whereas the region 
above this horizontal line corresponds to the free streaming region.  
In Case II, where the dark matter-electron coupling is due to electric dipole moment 
of the dark matter fermion, the free streaming of dark matter particles is allowed for 
the entire mass range and thus are not trapped inside the supernova. 
 In Fig. \ref{fig:Figure6}, we have plotted the lower bound on $\Lambda$ against the dark matter mass $m_\chi$ (considering $T_{SN}=30$ MeV, 
 $\rho_{SN}=3 \times 10^{14}$ gm/cc, $n_e=8.7 \times 10^{43} ~ \rm{m^{-3}}$).

\begin{figure}[htbp]
\includegraphics[width=10cm]{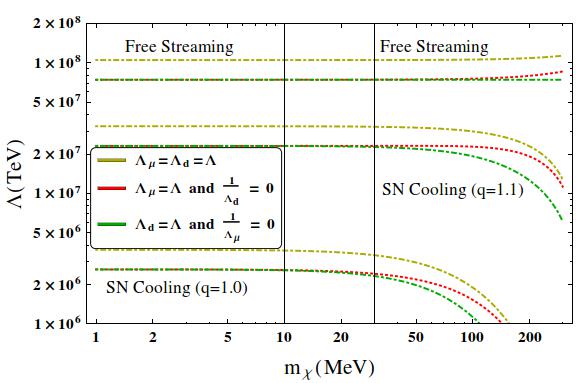}
\caption{ \it $\Lambda$ (in TeV) is plotted against $m_\chi$ (MeV). The lower set of curves follow from the SN1987A cooling in undeformed scenario(q=1.0), while the 
upper set of curves follows from the free streaming of dark matter fermions from the supernova core. The middle set of curves follows from the SN1987A cooling in q-deformed scenario(q=1.1). }
\protect\label{fig:Figure6}
\end{figure}

The upper(most) set of curves corresponds to that obtained using optical depth criteria (based on free streaming of dark matter fermion from the 
supernova core), whereas the lower(most) set of curves are obtained after applying 
the Raffelt's criteria on the SN1987A energy loss rate in undefromed case and the set of curves 
in the middle are obtained from the SN1987A cooling in q-deformed scenario(q=1.1). 
The region above both the curves (free streaming and SN cooling) is allowed for two different scenarios, q=1.0 and q=1.1.
correspond to $m_\chi = 10$ MeV and $30$ MeV. For 
a dark matter of mass $m_\chi = 30~\rm{MeV}$, the SN1987A energy loss rate 
gives a lower bound  $\Lambda(=\Lambda_\mu) = 2.4 \times 10^6~\rm{TeV}$ (lower curve) in undeformed scenario(q=1.0) and $\Lambda(=\Lambda_\mu) = 2.3 \times 10^7~\rm{TeV}$ (lower curve) in q-deformed scenario(q=1.1), whereas optical depth 
criterion (i.e. free streaming) fixes the lower bound on $\Lda(=\Lambda_\mu)$ at $7.42 \times 10^7~\rm{TeV}$ (lower curve). 


\section{Dependency of the effective scale $\Lambda$ on the supernova properties}
So far in our discussions we have not considered the variation of supernova properties like temperature($T_{SN}$), matter density($\rho_{SN}$), 
number density of electron($n_e$) which are relevant in the $ e \chi \rightarrow e \chi $ scattering process and on the variation of the lower bound 
on the effective scale $\Lambda$. The analysis was based on the following assumptions
\bea
T_{SN}=30~ \rm{MeV},~\rho_{SN}=3 \times 10^{14} ~ \rm{gm/cc},~n_e=8.7 \times 10^{43}~ m^{-3} \nonumber
\eea
where, we considered $T_{SN}$ is the average temperature of the supernova, $\rho_{SN}$ is the average matter density of the supernova and $n_e$ is the number density of the colliding electrons in the supernova (which are taking part in the $ e \chi \rightarrow e \chi $ scattering process).

 In the analysis below, we  propose the following
\begin{itemize}
\item[(i)] We can look for the fermionic DM emission pattern which is consistent with the Kamiokande \cite{Kamioka} and IMB \cite{IMB} data for neutrino emission due to supernova SN1987A explosion (which is more relevant as per our current understanding of supernova explosion mechanism). 
\item[(ii)] In another possibility, we can think about the annihilation process 
$ e^{-} e^{+}  \rightarrow \chi \bar{\chi} $ which is happening only at the outer $10\%$ of the proto-neutron star at some higher temperature (say $50 ~\rm{MeV}$) and some lower density (say $\rho_{SN}=10^{14}$ gm/cc) \cite{DHLP, Debades, Ellis, Lau, Sathees}. The justification of this second possible assumption is due to infall of matter particles in the accretion phase of the supernova, temperature is increasing at the outermost ($10\%$) region of the supernova core. Also due to the electron capture process by the protons, the central region of the proto-neutron star will be neutron-rich region and will not have much free electron-positron pair which can take part in the annihilation process.
\end{itemize} 
\subsection{Case I}
We can consider the supernova cooling phenomena due to the free streaming of dark matter 
fermions in both the phases, accretion phase and Kelvin-Helmholtz cooling phase. 
Due to infall of the matter objects in the accretion phase the temperature at the outermost 
region will increase than the average. The number density($n_e$) of 
the colliding electron (which takes part in $ e \chi \rightarrow e \chi $ scattering process) 
although appears to increase due to infall but actually not so. The infall 
boosts the production process of dark fermions(due to electron-positron annihilation:
$ e^- e^+ \rightarrow  \chi \overline{\chi}$ ) and the electron capture process by protons to 
create a neutron-rich core- together (annihilation 
and capture processes) result the number density($n_e$) to fall to a lower value.

\begin{figure}[htbp]
\includegraphics[width=10cm]{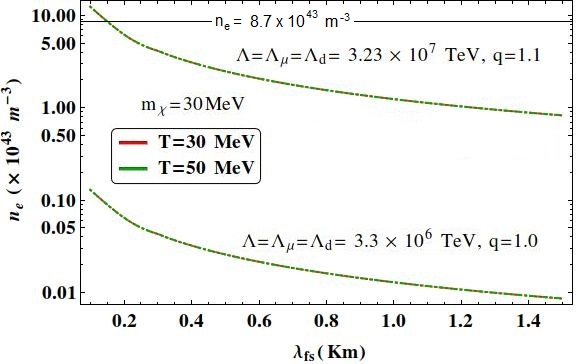}
\caption{\it The electron number density $n_e$ (in $m^{-3}$) is plotted against the electron's 
mean free path $\lambda_{fs}$ (in km). We have taken the dark matter fermion of mass 
$m_{\chi}=30~ \rm{MeV}$ and the value of the deformation parameter $q=1.0$(lower curve) and $q=1.1$(upper curve), respectively.}
\protect\label{fig:Figure7}
\end{figure}

 In Fig. \ref{fig:Figure7}, we have plotted the electron number density $n_e$ (in $m^{-3}$) 
against the mean free path $\lambda_{fs}$ (in km) for  $q=1.0$(undeformed scenario) and $q=1.1$(deformed 
scenario), respectively.

 We have considered the average supernova temperature $T=30 ~ \rm{MeV}$ and the number density 
 $n_e= 8.7 \times 10^{43}~ \rm{m^{-3}}$ \cite{Dreiner}. 
Considering the fact that the dark matter fermions, produced at a depth $r_0=0.9 R_c$, free stream 
out of the supernova and take away the released energy,  we find (from the optical depth criterion \cite{Dreiner}) 
that the minimum length of the mean free path for free streaming ($\lambda_{fs}$) is 
$\lambda^{min}_{fs} = 1.5~\rm{km}$. 

\begin{figure}[htbp] 
\includegraphics[width=10cm]{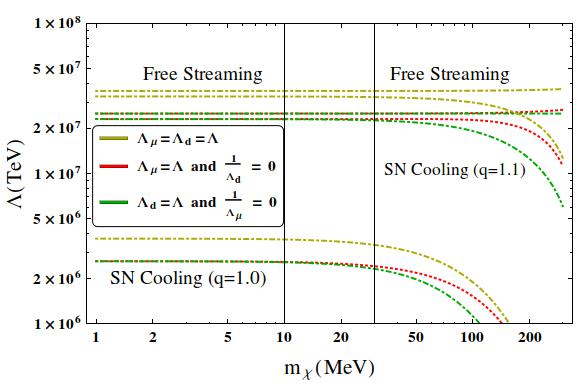}
\caption{\it The lower bound on $\Lambda$ (in TeV) (obtained from energy loss rate and the free streaming 
criteria) is plotted against $m_\chi$ (in MeV). The lower set of curves follows from 
the SN1987A cooling for q=1.0 and $T_{SN}=30$ MeV, while the 
curves in the above follow from the free streaming of dark matter fermions from the 
outermost region of the supernova core. The SNe cooling curves in q-deformed scenario(q=1.1) 
(the curves in the middle) are almost overlapping with the curves followed from free streaming.}
\protect\label{fig:Figure8}
\end{figure}

 From figure[\ref{fig:Figure7}], we see that if the value of the mean free 
 path $\sim 1.5 ~\rm{km}$, we need to choose $n_e \sim 10^{41}~ \rm{m^{-3}}$ in the undeformed 
 scenario(q=1.0) and $n_e \sim 10^{43}~ \rm{m^{-3}}$ in the q-deformed scenario(q=1.1), respectively. 
 Suppose at the outermost region the number density drops to $ \sim 10^{43}~ \rm{m^{-3}}$ 
 (from the average value $8.7 \times 10^{43}~ \rm{m^{-3}}$ ) and temperature rises to 
 $T=50~\rm{MeV}$ (above the average value  $T=30~\rm{MeV}$ ) which is consistent with existing supernova 
 simulation (matter density also decreases from its average value $ 3 \times 10^{14}$ gm/cc to  
 $10^{14}$ gm/cc) \cite{Debades} (we compared with the s23WH07 model as in that model mass of 
 the progenitor was considered to be $23M_{\odot}$ and progenitor of SN1987A had mass 
 $\sim 20M_{\odot}$). In Fig. \ref{fig:Figure8}, we have plotted the lower bound on $\Lambda$ against the dark matter 
 mass $m_\chi$ (obtained from the supernova energy loss rate
 and from the free streaming of the produced dark fermions from the 
 outermost region of the supernova core). The free streaming process is insensitive to the 
 temperature $T_{SN}$, we set $\rho_{SN}=3 \times 10^{14}$ gm/cc and 
 $n_e=1 \times 10^{43} ~ \rm{m^{-3}}$ in the free streaming case.

 The upper set of curves are obtained from the optical depth criteria (based on free streaming of dark matter fermion from the outermost part of the supernova core), 
 whereas the lower set of curves are obtained by applying the Raffelt's criteria on the
 SN1987A energy loss rate. The free streaming curves have shifted downwards from the one 
 obtained in figure[\ref{fig:Figure6}] and the SN cooling curves in q-deformed scenario(q=1.1) is almost merging with the set of 
 curves for free streaming.
 The two vertical lines correspond to $m_\chi = 10$ MeV and $30$ MeV, respectively. 
 For a dark matter of mass $m_\chi = 30~\rm{MeV}$, the SN1987A energy loss rate gives a lower 
 bound  $\Lambda(=\Lambda_\mu=\Lambda_d) = 3.3 \times 10^6~\rm{TeV}$ (upper curve) in the 
 undeformed scenario(q=1.0) and $\Lambda(=\Lambda_\mu=\Lambda_d) = 3.2 \times 10^7~\rm{TeV}$
 (upper curve) in the deformed scenario(q=1.1), 
whereas the optical depth criterion (i.e. free streaming) fixes the lower bound on 
$\Lda(=\Lambda_\mu=\Lambda_d)$ at $ 3.6 \times 10^7~\rm{TeV}$ (upper curve). This suggests 
almost all of the produced dark matter fermions can freely stream out of supernova to 
contribute to the supernova cooling with the fact that they are obeying q-deformed statistics: 
q=1.1 and with $n_e=1 \times 10^{43} ~ \rm{m^{-3}}$), 
which was not the case for $T=30$ MeV in undeformed scenario.

\begin{figure}[htbp] 
\includegraphics[width=7.5cm]{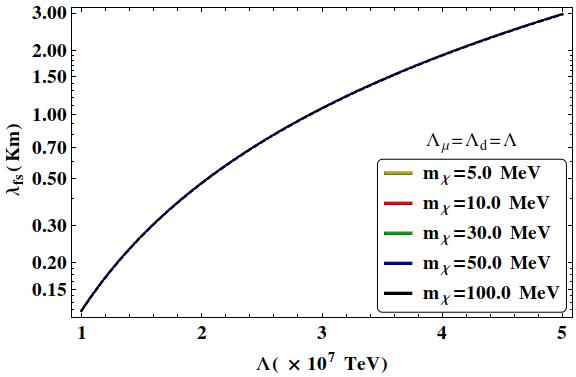} \hspace*{0.02in}
\includegraphics[width=7.5cm]{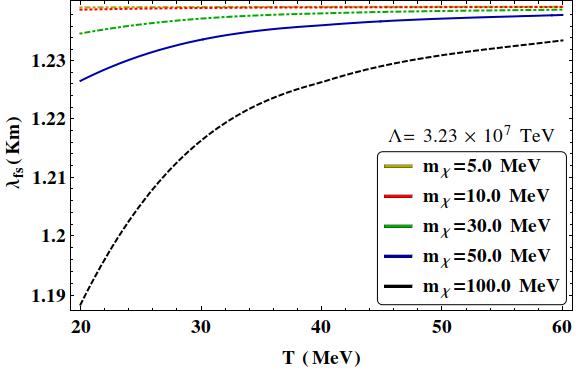}
\caption{\it {In the left figure $\lambda_{fs}$(in km) is plotted against $\Lambda$(in TeV) for 
$T=50$ MeV whereas in the right figure $\lambda$(in km) is plotted against T(in MeV) for 
$\Lambda_{\mu}=\Lambda_d=\Lambda=3.23 \times 10^7$ TeV (bound obtained from the SN cooling 
case in q-deformed scenario with q=1.1).}}
\protect\label{fig:Figure9}
\end{figure}

 For the crust temperature $T=50~\rm{MeV}$ and the number density of colliding electron 
 $n_e=10^{43}~ \rm{m^{-3}}$ we have plotted the mean free path($\lambda_{fs}$) as a function 
 of the effective scale($\Lambda$) and temperature($T$) in figure[\ref{fig:Figure9}] for 
 different values of dark matter fermion mass $m_{\chi}$.

 In figure[\ref{fig:Figure9}](left) we see that the mean free path $\lambda_{fs}$ increases 
 with the effective scale $\Lambda$ for different dark matter fermion mass. On the right, we see 
 that it first increases with temperature but eventually becomes constant with temperature $T$ and 
 they are different for different $m_{\chi}$ (GeV).
 
\begin{figure}[htbp]
\includegraphics[width=8.0cm]{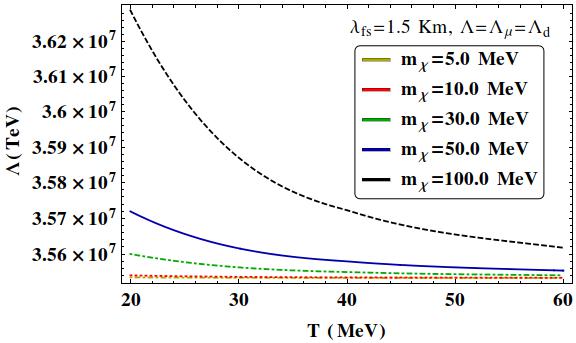} \hspace*{0.02in} \includegraphics[width=8.0cm]{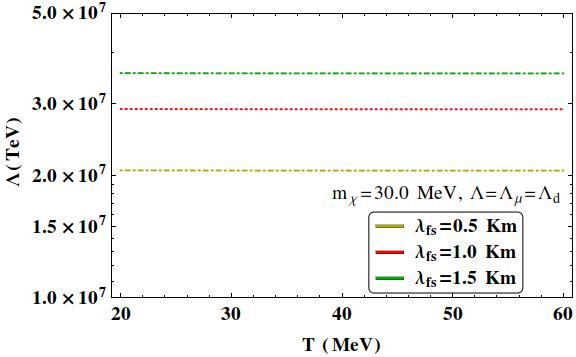}
\caption{ \it {$\Lambda=\Lambda_{\mu}=\Lambda_d$ (in TeV) is plotted against T (in MeV) where 
we have chosen $n_e=10^{43}~\rm{m^{-3}}$. The left figure corresponds
to the variation of $\Lambda$ with respect to $T$ for fixed mean free path $\lambda=1.5$ km 
and for different $m_{\chi}$ whereas the right figure shows the variation of $\Lambda$ with 
respect to $T$ for different mean free path $\lambda$ (in km) and for fixed $m_{\chi}=30$ MeV.}}
\protect\label{fig:Figure10}
\end{figure}

\begin{figure}[htbp], 
\includegraphics[width=9.6cm]{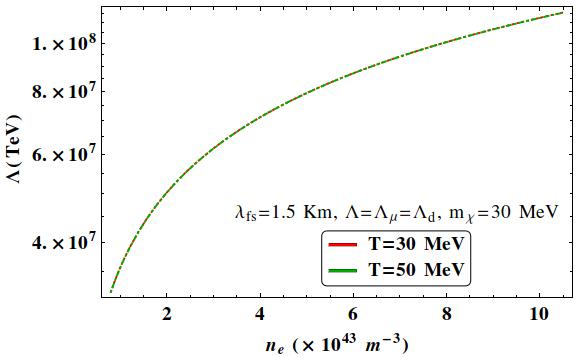}
\caption{ \it {The effective scale $\Lambda=\Lambda_{\mu}=\Lambda_d$ (in TeV) obtained for 
free streaming cases (using eqn.[\ref{mean_free_path}]) is plotted against $n_e$ 
(in $\rm{m^{-3}}$) for $\lambda_{fs}=1.5$ km, $m_{\chi}=30$ MeV. We find  
$\Lambda \sim 4 \times 10^7$ TeV for $n_e=10^{43}~ \rm{m^{-3}}$ and  
$\Lambda \sim 1 \times 10^8$ TeV for $n_e=8.7 \times 10^{43}~ \rm{m^{-3}}$.}}
\protect\label{fig:Figure11}
\end{figure}

 We next study the variation of the effective scale $\Lambda$ with temperature for different mass of dark matter fermions for a given $\lambda_{fs}$ and different values of mean free path for a 
 given $m_\chi$ and are shown in figure[\ref{fig:Figure10}]. 
From figure[\ref{fig:Figure10}] it is clear that the effective scale $\Lambda$ varies feebly 
with temperature, it remains almost constant.
Variation of the effective scale $\Lambda$ with the number density of colliding electrons 
$n_e$(which are taking part in the $ e \chi \rightarrow  e \chi $ scattering process) has 
been shown in figure[\ref{fig:Figure11}].
We find that $\Lambda$ increases considerably with the 
increment of the number density of colliding electrons $n_e$ (which are relevant in the 
discussion of the $ e \chi \rightarrow  e \chi $ scattering process). 
For $n_e=10^{43}~ \rm{m^{-3}}$, we find $\Lambda \sim 4 \times 10^7$ TeV and 
for $n_e=8.7 \times 10^{43}~ \rm{m^{-3}}$, $\Lambda \sim 1 \times 10^8$ TeV.

\subsection{Case II}
In section V, we considered the supernova core temperature $T_{SN}=30~\rm{MeV}$ and the core 
density as $\rho_{SN}=3 \times 10^{14}$ gm/cc. From the equations 
(\ref{eqn:energy loss rate final}, \ref{eqn:X}) it is clear that, the lower bound on the effective scale $\Lambda$ 
depends on the properties (temperature and density) of the supernova. As free streaming
is happening from the outermost $10\%$ of the supernova core, so it is justified to consider only 
crust (i.e., crust temperature and density) in the discussion of the supernova cooling phenomena as well. For core-collapse supernova, the crust temperature will be higher than the core whereas the 
density will fall from core to crust \cite{Debades, Sumiyoshi, Hagel}. In this section, we calculate 
the lower bound on $\Lambda$ in un-deformed ($q=1$) scenario using Raffelt's criteria and equation
(\ref{eqn:energy-loss-rate-undeformed}) considering temperature $T=50$ MeV and 
density $\rho=10^{14}$ gm/cc at the outermost sector (i.e. at $r = 0.9~R_c$, the crust) of the supernova core.
\begin{figure}[htbp], 
\includegraphics[width=10cm]{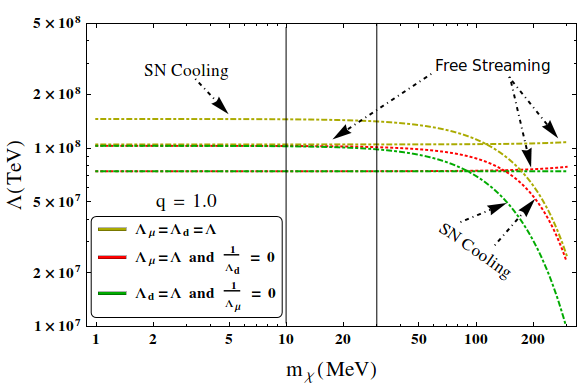}
\caption{ \it {The lower bound $\Lda=\Lda_{\mu}=\Lda_{d}$ (TeV) (in TeV) is plotted against the dark matter fermion mass $m_\chi$ (MeV) for two cases: 
(i) free streaming of dark matter fermions from the crust and (ii) from the Raffelt's criteria 
of SN1987A energy loss rate.}}
\protect\label{fig:Figure12}
\end{figure}

 In Fig. \ref{fig:Figure12}, we plot the lower bound on $\Lda(=\Lda_{\mu}=\Lda_{d})$ (TeV) as a function 
of $m_\chi$. The almost parallel set of curves (also parallel to x- axis) obtained using optical depth criteria
(based on the free streaming of dark matter fermion from the outermost sector of the supernova core), whereas 
the other set of curves are obtained after applying the Raffelt's criteria on the SN1987A energy loss rate (equation[\ref{eqn:energy-loss-rate-undeformed}]).
The region above each curve is allowed. The two vertical lines correspond to $m_\chi = 10$ MeV and $30$ MeV. For 
a dark matter of mass $m_\chi = 30~\rm{MeV}$, the SN1987A energy loss rate 
gives a lower bound  $\Lambda(=\Lambda_\mu=\Lambda_d) = 1.42 \times 10^8~\rm{TeV}$ (upper curve), 
whereas optical depth criterion (i.e. free streaming) fixes the lower bound on 
$\Lda(=\Lambda_\mu=\Lambda_d)$ at $1.05 \times 10^8~\rm{TeV}$ (upper curve). Note that the dark matter
fermions produced in the outermost sector(i.e. at $0.9 R_c$) can freely stream out to contribute in the 
supernova cooling phenomena which is not the case if they are produced at some inner region (i.e. at distance
 $r < 0.9 R_c$) of the supernova core. Even they are be copiously produced, they are not allowed to free stream 
(restricted by optical depth criterion) and hence can't contribute to the SN cooling as is seen from 
Figure[\ref{fig:Figure6}]. Some of them will transfer part of its energy back to the medium via scattering 
and as a result, the cooling process occurs slowly. 

 Below in Table 2 and Table 3, we summarize our result and make a comparative study  between two cases as discussed in section V and 
section VI:
\begin{center}
Table 2
\end{center}
\begin{center}
\begin{tabular}{|c|c|c|c|}
\hline
SN Properties & $m_\chi~(\rm{MeV})$ & \multicolumn{2}{c|}{$\Lda=\Lda_\mu=\Lda_d~\rm{(TeV)}$}    \\
\cline{3-4}
  &   &  Free Streaming & SN Cooling \\
\hline
 $T_{SN}=30$ MeV,  & 10 & $1.05\times 10^8$ & $3.66\times 10^6$ \\ 

\cline{2-4}
  $\rho=3 \times 10^{14}$ gm/cc & 30 & $1.05\times 10^8$ & $3.34\times 10^6$ \\ 

\hline
  $T_{SN}=50$ MeV,  & 10 & $1.05\times 10^8$ & $1.45\times 10^8$ \\

\cline{2-4}
  $\rho=10^{14}$ gm/cc & 30 & $1.05\times 10^8$ & $1.42\times 10^8$ \\

 \hline

\end{tabular}
\end{center}
\noindent {\it Table 2: The lower bound on the effective scale $\Lda=\Lda_{\mu}=\Lda_{d}$ (TeV) (obtained from Raffelt's criteria and optical depth criterion) are shown for different 
dark matter mass $m_\chi$ (MeV) in the undeformed scenario for different values of SN temperature 
$T(=T_{SN})$ and matter density $\rho_{SN}$.}

 \noindent From Table 2, we see that as $m_\chi$ increases from $10~\rm{MeV}$ to $30~\rm{MeV}$, the bound on 
$\Lda$ decreases (follows from SN energy loss rate) whereas for free streaming case (optical depth criterion)
$\Lda$ remains constant. We also see from Table 2 that the bound on $\Lambda$ change as a function of the
supernova properties (i.e. it's temperature and density). In Table 3, we have shown the 
lower bound $\Lda=\Lda_{\mu}=\Lda_{d}$ (TeV) (obtained from Raffelt's criteria and optical 
depth criterion) for different dark matter mass $m_\chi$ (MeV) in the undeformed scenario and q-deformed scenario for different 
values of supernova crust temperature $T$  and different values of the electron density $n_e$. 
\begin{center}
Table 3
\end{center}
\vspace*{-0.25in}
\begin{center}
\begin{tabular}{|c|c|c|c|c|}
\hline
SN Properties & $m_\chi~(\rm{MeV})$ & \multicolumn{3}{c|}{$\Lda=\Lda_\mu=\Lda_d~\rm{(TeV)}$}  \\
\cline{3-5}
  &   &  Free Streaming & SN Cooling(q=1.0) & SN Cooling(q=1.1) \\
\hline
 $T=30$ MeV,  & 10 & $1.05\times 10^8$ & $3.66\times 10^6$ & $3.23 \times 10^7$ \\ 

\cline{2-5}
  $n_e=8.7 \times 10^{43}~\rm{m^{-3}}$ & 30 & $1.05\times 10^8$ & $3.34\times 10^6$ & $3.23 \times 10^7$ \\ 

\hline
  $T=50$ MeV,  & 10 & $3.6 \times 10^7$ & $3.66\times 10^6$ & $3.23 \times 10^7$ \\

\cline{2-5}
 $n_e=1 \times 10^{43}~\rm{m^{-3}}$ & 30 & $3.6\times 10^7$ & $3.34\times 10^6$ & $3.23 \times 10^7$ \\

 \hline

\end{tabular}
\end{center}
\noindent {\it Table 3: The lower bound on the effective scale $\Lda=\Lda_{\mu}=\Lda_{d}$ (TeV) 
(obtained from Raffelt's criteria and optical depth criterion) are shown for different dark 
matter mass $m_\chi$ (MeV) in the undeformed scenario and q-deformed scenario for different 
values of SN crust temperature $T$. We have considered the variation of the temperature 
and the number density of colliding electrons $n_e$ only at the crust level.}

 If we consider different phases of a typical core-collapse supernova like the accretion and Kelvin-Helmholtz cooling phase, then the bounds will 
change. We can compare the bounds with the knowledge of temperature and density of supernova in those 
phases.

\section{Conclusion}
The dark matter fermions, pair produced in electron-positron collision $e^+ e^- \to \chi \overline{\chi}$ inside 
the supernova core, can take away the energy released in the supernova SN1987A explosion. 
Working within the formalism of $q$-deformed 
statistics (as the core supernovae temperature is fluctuating with the average value $T_{SN} = 30~\rm{MeV}$) and using the Raffelt's criterion on the 
emissivity for any new channel $\dot{\varepsilon}(e^+ e^- \to \chi \overline{\chi}) \le 10^{19}~{erg~g^{-1}s^{-1}}$, we find that as the deformation parameter $q$ changes 
from $1.0$ (undeformed scenario) to $1.1$(deformed scenario), the lower bound on the scale $\Lda$ of the dark 
matter effective theory varies from $3.3\times 10^6$ TeV to $3.2 \times 10^7$ TeV for a dark matter fermion of mass $m_\chi = 30~\rm{MeV}$. 
Using the optical depth criteria on the free streaming of the dark matter fermion, we find the lower 
bound on $\Lambda \sim 10^{8}~\rm{TeV}$ for  $m_\chi = 30~\rm{MeV}$. 
In a scenerio,where the dark matter fermions are pair produced in electron-positron annihilation in the 
outermost sector of the supernova core (with radius $0.9 R_c \le r \le R_c$  where $R_c (=10~\rm{km})$ being 
the supernova core radius or the radius of proto-neutron star), we find that the bound on $\Lambda$ 
($\sim 3 \times 10^7$ TeV) obtained from SN cooling criteria (Raffelt's criteria) is comparable  with the bound  obtained from free streaming (optical depth criterion) for light fermion dark matter of mass 
 $m_{\chi}=10 - 30$ MeV. In a nutshell, all the dark matter fermions produced in the 
 outermost sector (i.e. at the crust) can freely stream out to contribute to the supernova cooling phenomena 
 which is not the case if they are producing in some inner region than the crust.


\section{Acknowledgments}
\noindent The authors would like to thank Dr. Partha Konar,PRL,Ahmedabad and Prof. Debasish Majumdar, 
SINP, Kolkata and Prof. Tirthankar Roychaudhury for discussions. This work is partially funded by the Board of Research in Nuclear Science, Department of Atomic Energy,
Government of India, Grant No. 2011/37P/08/BRNS.
\section{Feynman rules}
\label{app:feyn}
Process: $e^- e^+ \stackrel{\gamma}{\longrightarrow}  \chi \overline{\chi}$: 

{  $e^-~e^+ \to \gamma$ ~vertex: } $ie \gamma^\mu$ 

{  $\gamma \to \chi~\overline{\chi}$ ~vertex: } $i\left( \mu_\chi \sigma^{\mu \nu} q_\nu 
                                               + d_\chi \sigma^{\mu \nu} q_\nu \gamma^5 \right)$ 

\section{APPENDIX:}
\subsection*{From q-deformed statistics to undeformed scenario}
 In general, the distribution function for the q-deformed statistics is \cite{Tsallis}
\bea
D_i=\left(1+\frac{b}{\tau}(E_i-\mu_i)\right)^{\tau}+1
\eea 
with $b=\frac{\beta_0}{4-3q}$, $\beta_0=\frac{1}{k_B T}$ (we work in the unit $k_B=1$) and $\tau=\frac{1}{q-1}$

 In terms of the dimensionless quantity $x_i=\frac{E_i}{T}$
\bea
D_i=\left(1+b (q-1)(T x_i-\mu_i)\right)^{\frac{1}{q-1}}+1
\eea 
 
 Now replacing $q-1$ by $m$, ($m \rightarrow 0$ as $q \rightarrow 1 $)
\bea
\left(1+b (q-1)(T x_i-\mu_i)\right)^{\frac{1}{q-1}}=\left(1+b m(T x_i-\mu_i)\right)^{\frac{1}{m}}=y \mathrm{(say)}
\eea  
Now
\bea
\lim_{m\to 0} y &=& \lim_{m\to 0} \left(1+b m(T x_i-\mu_i)\right)^{\frac{1}{m}} \nonumber \\
\implies \lim_{m\to 0} \ln y &=& \lim_{m\to 0} {\frac{1}{m}} \ln \left(1+b m(T x_i-\mu_i)\right)
\nonumber \\
&=& \lim_{m\to 0} \frac{1}{1+b m(T x_i-\mu_i)} b (T x_i-\mu_i)
\nonumber \\
&=& b (T x_i-\mu_i) \nonumber
\eea
 Also for $q \rightarrow 1 $, we find $b(=\frac{\beta_0}{4-3q})=\beta_0=\frac{1}{k_B T}=\frac{1}{T}$. So
we find 
 \bea
 \lim_{q\to 1} \ln y &=& \beta_0 (T x_i-\mu_i) 
 \nonumber \\
\implies \lim_{q\to 1} y &=& \exp \left[\frac{1}{T} (T x_i-\mu_i) \right]
 \nonumber \\
 &=& \exp \left[x_i- \frac{\mu_i}{T} \right] \nonumber
 \eea
 Clearly, in the undeformed scenario (i.e. $q=1 $)
 \bea
 \lim_{q\to 1} D_i &=& \lim_{q\to 1} y +1  \nonumber \\
  &=& \exp \left[x_i- \frac{\mu_i}{T} \right] +1 ~~[Proved]
 \eea


\end{document}